# The impact of private sector credit on income inequalities in European Union (15 member states)

**Ionuţ JIANU**
The Bucharest University of Economic Studies, Romania
ionutjianu91@yahoo.com

**Abstract.** *This paper aims to provide a comprehensive analysis on the income inequalities recorded in the EU-15 in the 1995-2014 period and to estimate the impact of private sector credit on income disparities. In order to estimate the impact, I used the panel data technique with 15 cross-sections for the first 15 Member States of the European Union, applying generalized error correction model.*

**Keywords:** credit, error correction model, European Union, income inequality, panel.

**JEL Classification:** C33, D63, E51.

## 1. Introduction

The financial crisis has questioned the financial sector's role in the development and growth process and its impact on income distribution. National governments considered that the banking system is too big to fail and, finally, guaranteed for the banking sector debt with government bonds to save it, subsequently transforming private debt into public debt. Indeed, the world could not pass through this stage without as policymakers choosing a compromise between the financial and the social one. Most of the research papers developed in this area have demonstrated that national governments opted for a compromise that has deepened the social imbalances in the world. The theory says that an exaggerated level of lending has the ability to increase the economy's vulnerability to shocks and to cause imbalances on the distribution of credit, which may affect the income gap, as individuals with low income will not benefit from the same opportunities as those of individuals with high income.

In 2014, the private sector credit - expressed as a share of gross domestic product (GDP) - from the first 15 European Union member states (EU-15) increased by 58.22% compared to the level recorded in 1995, while Gini index (relevant indicator for income inequality) fell by 0.32% on the same timeframe. Although, on average, income inequality is considerably lower in 2014 than it level from 1995, this indicator had an extremely heterogeneous evolution in the EU-15 Member States, its evolution being influenced by other factors, which were integrated into the research.

The trend of income inequality is considered a result of the indicators evolution related to economic growth, social, education and health spending, financialisation, corruption and national institutional features, unemployment, investment, tax burden, economic openness, technological progress, price level, as well as energy capture by human activity and information processing.

The motivation for choosing this theme lies in the academic interest for income inequality area and its constant presence in the agenda of economic debates, respectively the social discontent against the capitalist system reforms.

The main goal of the paper is to estimate the impact of private sector lending on income inequalities, it following to be achieved by reaching three specific objectives:
- identifying the impact of private sector lending and other relevant factors that have the ability to impact the income inequality;
- testing the Gauss-Markov theorem hypotheses;
- performing the economic analysis of the obtained results.

## 2. Literature review

The results of researches whose goal consisted in estimating the impact of financial development on income inequality, are quite ambiguous and often contradictory. Galor and Moav (2004) proved that credit growth leads to a decrease in income inequalities level, given the fact that credit growth give the opportunity to the lower class to borrow and start new projects, which can reduce the income gap. On the other hand, Rajan and Zingales (2003) showed that in countries with weak institutions, there is a positive relationship between financial development and income inequality, considering that individuals with high incomes have a privileged position in terms of access to finance. According to Law et al. (2014), financial development could represent a driver force of income inequalities reduction, in the case of prevailing strong institutions, which allows individuals of the lower-class to invest in human capital.

Clarke et al. (2006) and Beck et al. (2007) identified a positive relationship between financial development and the gap between incomes, while Jauch and Watzka (2012) found a positive impact of credit growth on income inequality, using a panel data with fixed effects for 138 countries.

Jaumotte et al. (2013) analysed the concept of income inequality as a consequence of financial and trade globalization. The analysis included private lending as a share of GDP on the control variables list, the corresponding coefficient for this variable being positive and statistically significant. Dabla-Norris et al. (2015) analysed the financial development and its overwhelming influence on increasing income disparities in the 1980-2012 period in 97 states and have reached similar conclusions.

According to Piketty (2014), financial expansion can lead to an increase in income inequality because of its statistical association with wealth increase, the last one being distributed more unevenly than the incomes. Also, Li and Yu (2014) estimated the impact of lending as a share of GDP on income inequality (expressed in Gini indicator), identifying a significant and positive coefficient of it. Moreover, Denk and Cournède (2015) analysed a sample of 33 OECD countries and proved that the existence of a high degree of financialisation coincides with a situation where income disparities prevailing the national state. According to them, the expansion of financial creates the premises of income inequalities increase, given that economic entities with high profits can receive larger loans than the ones of individuals or economic entities with low incomes, leading to a higher return on investment for high profits economic operators, which leads to wage increases. Furthermore, de Haan and Sturm (2016) identified a positive impact of financial development, financial liberalization and banking crises on income inequality, the effect being conditioned by the features of political institutions.

The selection of the private sector credit as a relevant variable to capture the impact of finances on income inequality came from the hypothesis demonstrated by Naceur and Zhang (2016), according to which its effect is manifested mainly through the banking sector, but not through the stock market capitalization.

Testing the relationship between credit and income inequality has raised new questions concerning the manifestation of the endogeneity between inequality and lending. Most of the existing research in this field, such as those of Rajan (2010) or Kumhof (2015) showed that loan growth may be a consequence of the growing gap between incomes, taking into consideration the fact that individuals who get low incomes start borrowing to avoid consumerist disparities against the upper class. Also, Gu and Huang (2014) confirmed this hypothesis. Van Treeck (2014) found that the financial crisis in the United States (2007-2008) was caused by the increase of the income gap, following the same economic foundation mentioned above. On the other hand, the literature review provides conflicting views of the authors too. Atkinson and Morelli (2011), respectively Bordo and Meissner (2012) did not identify any significant impact of income inequality on the credit cycle. However, the literature is not conclusive on the assumption of this hypothesis. Most of the papers from this field examined the relationship between financial liberalization and income inequality, as well as that of Claessens and Perotti (2007) saying that extractive institutions continue to favor a rent-seeking behavior.

## 3. Methodology

The main objective of the paper is to estimate the impact of private sector lending on income inequalities in the EU-15. In this context, I used the quantitative approach, deductive method and certain specific econometric techniques.

In order to investigate the stated objective, I estimated the impact of private sector lending growth and other control variables (listed in Annex 1) on income inequalities – in Eviews 9.0 software – using panel technique and generalized error correction model, implicitly the Generalized Least Squares method (the estimation being weighted with the Cross-Section SUR option).

Data series were extracted on yearly basis, covering the 1995-2014 period for the first 15 Member States of the European Union. I chose this representation of the panel data, because these countries were member states of the European Union on the entire period of analysis, thereby, the research not being affected by the adhesion of other countries to the union. Statistical database used did not cover the entire time horizon analysed for the Gini coefficient and lending to the private sector (expressed as a percentage of GDP), which made it necessary running the linear interpolation tool of Eviews, in order to estimate the missing data from the analysis (Annex 2 and Annex 3).

Following the verification of the stationarity and of the residuals, it has resulted the use of error correction model as the most accurate method for estimation. In order to confirm the econometric method used, I checked the stationarity test using the Summary window, a technique that provides results for each of the 5 tests applied for unit root[1] assumption and Hadri, in exceptional cases. However, the impact analysis of the exogenous variables on a single endogenous variables and the non-stationary nature of the initial variables (becoming stationary after performing the first difference) indicated selecting error correction model as the most appropriate estimation method, eliminating the vector error correction model from the possible alternatives. The selected method requires adding the error term lagged by 1 time-series frequency (resulting from the estimated model using the initial variables), its specific coefficient having the role to estimate the speed of adjustment. Error correction model can be applied only if the error term is stationary, a situation that indicates a long-term relationship between regressors and regressand.

Even if the panel is considered a more effective technique because it increases the number of degrees of freedom and the estimation efficiency, it can also bring new challenges, regarding the autocorrelation between cross-sections. In this respect, I used Cross-Section SUR option to correct, ex-ante, the possible inconveniences of the model.

After including the interpolated data, the first differences of the variables and the appropriate lag in the model, it has resulted 270 observations of the 300 initial observations (15 cross-sections). Regarding the selection of the appropriate lag, I used the Schwarz information criterion for each of the first four lags (Schwarz has the smallest value for lag 1, concluding that choosing lag 1 is appropriate).

Identifying the optimal lag, the differentiation level, and the estimation method, has made it possible the estimation of the following model:

$$D(Gini) = \alpha_0 + \alpha_1 D(Gini_{t-1}) + \alpha_2 D(creditp_{t-1}) + \alpha_3 D(he_{t-1}) +$$
$$+ \alpha_4 D(ed_{t-1}) + \alpha_5 D(un_{t-1}) + \alpha_6 D(tax_{t-1}) + \alpha_7 D(CPI) +$$
$$+ \alpha_8 D(openness) + \alpha_9 D(gfcf) + \alpha_{10}(UT_{t-1}) + \varepsilon_t \qquad (1)$$

where:

$D(Gini)$ and $D(Gini_{t-1})$ catch the first difference of Gini coefficient and the first difference of its auto-regressive term, while $D(creditp_{t-1})$, $D(he_{t-1})$, $D(ed_{t-1})$, $D(un_{t-1})$, $D(tax_{t-1})$ represents the first difference of the following variables lagged by one year: private sector credit as a share of GDP, government health expenditures as a share of GDP, government education expenditures as a share of GDP, unemployment rate, respectively total tax burden as a share of GDP (including imputed social security contributions). On the other hand, $D(CPI)$, $D(openness)$ and $D(gfcf)$ correspond to the dynamic of the corruption perception index, trade openness and gross fixed capital formation. Finally, the last component of the model is $UT_{t-1}$ – the error term laged by one year – respectively $\varepsilon_t$, this series being estimated in order to indicate the residuals distribution.

After estimating the model, I checked the assumptions of the Gauss-Markov theorem to confirm or reject that the estimators are best linear unbiased estimators. Thereby, I used the verification of the following hypotheses by the methods mentioned-below at a significance threshold of 5%:
- linearity of the model;

- the confirmation of the significance of the parameters and non-zero dispersion for each regressor (T-test and standard error);
- the existence of a number of observations greater than the number of coefficients;
- the absence of the multicolinearity, the test being performed by using Klein criterion - the absence of the multicolinearity is confirmed when the correlation coefficient between two exogenous variables is less than the coefficient of determination;
- the absence of correlation between regressors and residuals (Pearson correlation);
- the confirmation of the errors features, according to which their average is null and the residuals are normally distributed (Jarque-Bera test and histogram of the residuals);
- the confirmation of the errors features, according to which their variance is constant (the existence of the homoskedasticity – White test);
- the absence of cross-sections dependence (Breusch-Pagan and Pesaran test);
- the confirmation of the hypothesis, according to which conditional average of the errors is null (Residuals plot).

## 4. Results and interpretations

This section examines the impact of private sector credit on the evolution of income inequality, as well as investigates the validity of the model.

Initially, I checked the stationarity of the variables using the tests mentioned in methodology to identify the appropriate procedure for estimating the model. The tests performed indicated the non-stationary character of the variables, these becoming stationary after processing the first difference (Table 1).

**Table 1.** *Stationarity tests – Summary window*

| Variable | Number of tests rejecting I(0) unit root hypothesis (α = 5%) | | | Number of tests rejecting I(1) unit root hypothesis (α = 5%) | | |
|---|---|---|---|---|---|---|
| | Individual intercept | Individual intercept and trend | None | Individual intercept | Individual intercept and trend | None |
| Gini* | 1 of 4 | 5 of 5 | 0 of 3 | 4 of 4 | 5 of 5 | 3 of 3 |
| Gini(-1)* | 1 of 4 | 4 of 5 | 0 of 3 | 4 of 4 | 5 of 5 | 3 of 3 |
| creditp(-1)* | 3 of 4 | 2 of 5 | 0 of 3 | 4 of 4 | 4 of 5 | 3 of 3 |
| he(-1) | 0 of 4 | 2 of 5 | 0 of 3 | 4 of 4 | 5 of 5 | 3 of 3 |
| ed(-1) | 1 of 4 | 2 of 5 | 0 of 3 | 4 of 4 | 5 of 5 | 3 of 3 |
| un(-1) | 2 of 4 | 0 of 5 | 0 of 3 | 4 of 4 | 5 of 5 | 3 of 3 |
| tax(-1) | 0 of 4 | 0 of 5 | 0 of 3 | 4 of 4 | 5 of 5 | 3 of 3 |
| CPI* | 2 of 4 | 3 of 5 | 0 of 3 | 4 of 4 | 5 of 5 | 3 of 3 |
| openness* | 1 of 4 | 5 of 5 | 0 of 3 | 4 of 4 | 5 of 5 | 3 of 3 |
| gfcf | 0 of 4 | 1 of 5 | 0 of 3 | 4 of 4 | 5 of 5 | 3 of 3 |

*Hadri test rejects the stationarity hypothesis at level and accepts it after processing the first difference.
**Source:** Own calculations using Eviews 9.0

The use of error correction model involves adding in the regression the variables specific to first differences, as well as of the error term, this process being conditioned of its stationary character. In this respect, I checked and confirmed the error term stationarity using the same procedure mentioned above (Figure 1).

**Figure 1.** *Error term stationarity*

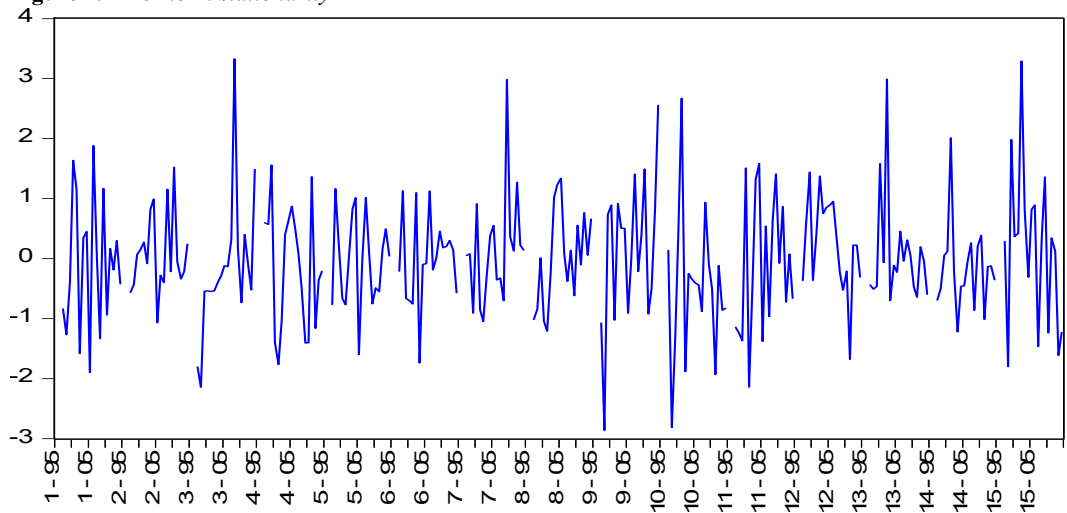

**Source:** Own calculations using Eviews 9.0

Next, I estimated the model by the Generalized Least Squares method (Cross-Section SUR option), the obtained results being attached in the Annex 4. Their analysis indicates a coefficient of determination of 0.8093, confirming the linearity of the regression and the proper selection of the regressors, given that the exogenous variables evolution explains 80.93% of the fluctuation of the regressand. Also, the probability of the F-test (0%) confirmed the validity of the model in statistical terms. Following that, I analysed the impact of the regressors, respecting the „ceteris paribus" assumption.

According to the results attached in Annex 4, the increase of autoregresive term with 1 deviation point led to an increase in the dynamic of the Gini coefficient with 0.899 points. The impact can be caused by the ability of the savings to generate new additional revenues for the population that records high-incomes, which may increase the spread between wage incomes.

Regarding the subject of interest of the paper, the model has estimated an impact of 0.005 points on the dynamics of Gini index at an increase with 1 percentage point of credit to private sector as a share of GDP – lagged by one year in the EU-15. This effect results from approving, in particular, of loans to companies with high turnovers, this type of loan being less risky. Subsequently, these economic entities increase their profits by adopting and applying certain development strategies (as a result of accessing the credited amounts), which can lead to an increase in wages, increasing the disparities between employees working in a company with low profits and low access to credit and those who operate in a company with high profits and high access to credit. Another explanation lies in the effect of the credit growth on the rise of asset prices, implicitly on investors wealth.

The control variables used have contributed significantly to the fluctuation of income inequality. The increase by 1 percentage point in the dynamics of government spending on health (as % of GDP) and government spending on education (as % of GDP) – variables lagged by one year – reduced the dynamic of Gini coefficient with 0.104 and 0.139 points, the explanation of the phenomenon being given by the theory of human capital. On the other hand, the 1 percentage point growth in the dynamics of unemployment rate from the previous year has led to an increase in the dynamic of income disparities from the current year with 0.063 points, this relationship being argued by the impact of reducing the number of employees in the economy (implicitly that of the transition from wages to income from unemployment benefits - which are lower than wage incomes) on income gap. Regarding the impact of the tax burden on Gini index, an increase in its dynamics (variable lagged by 1 year) by 1 percentage point grew the spread between incomes with 0.045 points.

The variables included in the model that are impacting the Gini coefficient on short term (corresponding to the situation where there is no lag) are the level of corruption, openness of the economy and gross fixed capital formation. In this respect, the increase of the dynamic of corruption perception index by 1 deviation point (which shows a decrease in the level of corruption, considering the reverse scaling of the indicator) had an effect of –0.029 on the dynamics of Gini index, influence that could be argued by multiple channels of impact, such as: the impact of corruption on economic growth and, implicitly on income inequality, the impact of corruption on tax evasion, ineffective tax administration and on the deductions through social groups benefits disproportionately, the effect of corruption on reducing the efficiency of social programs or the impact of corporate lobbies on the policymaking process, due to the high concentration of property assets in the economy. Regarding the impact of the openness of the economy on the income inequalities, the increase of the dynamic of the trade openness by 1 percentage point led to a decrease of income inequality dynamics with 0.019 points, the effect having the ability to manifest through the influence of trade on economic growth channel, which reduce the Gini coefficient, as well as that of the impact of trade openness on the expansion of certain market sectors.

On the other hand, the increase of gross fixed capital formation dynamic with 1 percentage point increase the dynamic of Gini index with 0.081 deviation points.

If all the variables are constant, the dynamic of the gap between incomes feels a growth with 0.060 points. Taking into account the long-run relationship between regressors and regressand, I have found that the annual speed of adjustment of the disequilibriums was 1.122%, this coefficient being statistically valid, given that it satisfies the condition of the negative sign and its statistically significant coefficient.

The estimated coefficients for each exogenous variable are significant at a significance threshold of 1%, excepting the first differences of government spending on education coefficient, a variable that is statistically significant for the 5% threshold, but not for the 1%. Also, the standard errors of the estimators is non-zero, but close to 0, which confirms the second hypothesis of the Gauss-Markov theorem, while relatively high population size (the third hypothesis) creates the premises for a proper representation of the residuals.

Table 2 shows that there is no multicolinearity between the regressors used in the analysis, taking into account that the coefficient of determination is higher than the statistical correlation coefficient between the exogenous variables.

**Table 2.** *Klein criterion – testing multicollinearity*

| Correlation matrix of 1st difference | Gini* | cred* | he* | ed* | un* | tax* | CPI | open | gfcf |
|---|---|---|---|---|---|---|---|---|---|
| Gini* | 1.00 | 0.03 | 0.01 | 0.02 | -0.03 | 0.00 | 0.02 | -0.06 | 0.00 |
| cred* | 0.03 | 1.00 | 0.14 | 0.11 | -0.09 | -0.02 | -0.04 | -0.06 | -0.08 |
| he* | 0.01 | 0.14 | 1.00 | 0.45 | 0.14 | -0.18 | 0.05 | -0.08 | -0.24 |
| ed* | 0.02 | 0.11 | 0.45 | 1.00 | 0.23 | -0.03 | 0.04 | -0.07 | -0.23 |
| un* | -0.03 | -0.09 | 0.14 | 0.23 | 1.00 | 0.01 | 0.02 | 0.14 | -0.40 |
| tax* | 0.00 | -0.02 | -0.18 | -0.03 | 0.01 | 1.00 | 0.05 | 0.02 | 0.18 |
| CPI | 0.02 | -0.04 | 0.05 | 0.04 | 0.02 | 0.05 | 1.00 | 0.07 | 0.03 |
| open | -0.06 | -0.06 | -0.08 | 0.07 | 0.14 | 0.02 | 0.07 | 1.00 | 0.16 |
| gfcf | 0.00 | -0.08 | -0.24 | -0.23 | -0.40 | 0.18 | 0.03 | 0.16 | 1.00 |

R-squared = 0.8093; open is openness and cred is creditp; * are the regressors lagged by 1 year.
**Source:** Own calculations using Eviews 9.0

On the other hand, the Table 3 confirms the hypothesis concerning the absence of correlation between regressors and residuals, which validates other two hypothesis of Gauss-Markov theorem.

**Table 3.** *Correlation matrix between residuals and regressors*

| Correlation matrix* | Gini** | cred** | he** | ed** | un** | tax** | CPI | open | gfcf |
|---|---|---|---|---|---|---|---|---|---|
| Residuals | -0.06 | 0.01 | 0.01 | -0.02 | -0.02 | 0.01 | -0.01 | 0.00 | -0.01 |

*Regressors are expressed in 1st difference form; ** are the regressors lagged by 1 year.
**Source:** Own calculations using Eviews 9.0

According to the result of Jarque-Bera test (Table 4), there are no arguments to reject the hypothesis related to normal distribution of errors, since its probability-value (0.186) is above the threshold of 5% used. At the same time, the average of the residuals is null, making it possible the investigation of the following hypotheses. After using White test, I have accepted the hypothesis that the model is homoskedastic, given that the product of the number of observations and R-squared (218.515) is lower than Chi-square statistic (296.466) for 258 degrees of freedom. Also, Breusch-Pagan LM (1.000) and Pesaran CD (0.863) tests confirm the absence of cross-sections dependence, since their probability is higher than the significance threshold of 5%.

**Table 4.** *Residuals hypothesis*

| Hypothesis | Method | Result |
|---|---|---|
| Normal distribution of residuals | Jarque-Bera | 0.186 |
| Homoskedasticity | White-test | 218.515 |
| Cross-section dependence absence | Breusch-Pagan LM | 1.000 |
| Cross-section dependence absence | Pesaran CD | 0.863 |

**Source:** Own calculations using Eviews 9.0

Regarding the verification of the zero conditional mean of the errors, I used the Residuals plot to check the constancy of residuals. Figure 2 demonstrates the constant evolution of the residuals (blue line), around 0. Thereby, I have accepted the hypothesis related to zero conditional mean of the errors and afterwards, I have validated the model - the estimators being characterised by maximum verisimilitude.

**Figure 2.** *Actual, fitted, residual plot*

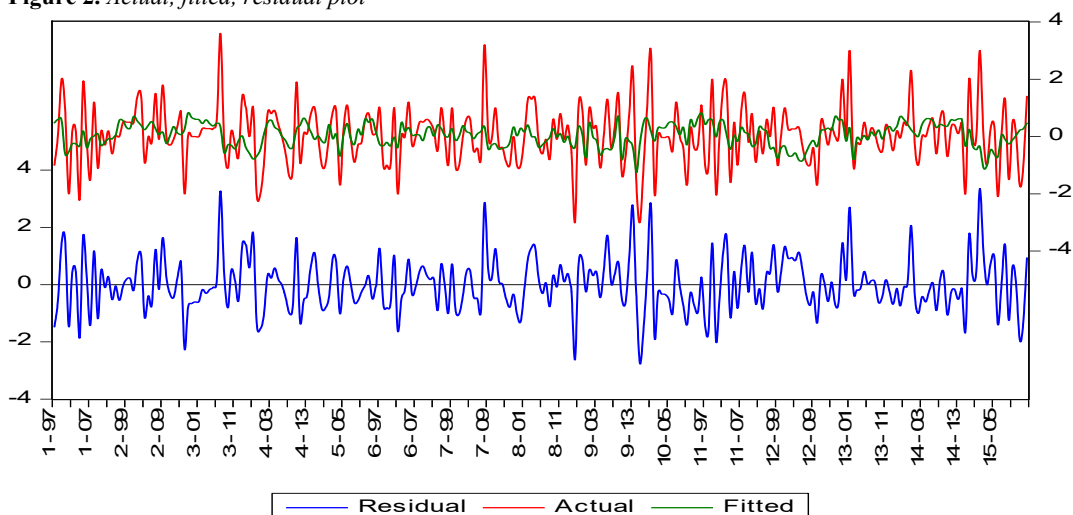

**Source:** Own calculations using Eviews 9.0

In other words, the model has fulfilled all the conditions for a correct statistical representation and its expected effects can be considered very close to the current reality.

## 5. Conclusions

This paper estimated a positive impact of private sector credit growth on income inequality, this being caused by the existence of disequilibriums in the distribution of the credit. The coefficient is statistically significant and the model meets all Gauss-Markov theorem assumptions, necessary to the validation of the maximum verisimilitude of the estimators. Moreover, there is a long-run relationship between the exogenous variables of the model and the Gini coefficient. Also, this research demonstrated that the income inequality has a historical cause. The proper conduct of macro-prudential policy can provide a solution in terms of moderating the impact of excessive lending to the private sector on the income inequalities.

**Note**

[1] See Levin, Lin and Chu (i), Breitung (ii), Im, Pesaran and Shin (iii), ADF-Fisher (iv), PP-Fisher (v).

**Annex 1.** *Data used*

| Indicators | Source |
|---|---|
| GINI coefficient | Eurostat and OECD |
| Private sector credit (as a % of GDP) | World Bank |
| Health government expenditures (as a % of GDP) | Eurostat |
| Education government expenditures (as a % of GDP) | Eurostat |
| Unemployment rate (%) | International Monetary Fund |
| Total tax burden including imputed social security contributions - total economy (as a % of GDP) | AMECO |
| Corruption perception index | Transparency International |
| Total trade (as a % of GDP) - openness index | Eurostat |
| Fixed gross capital formation (as a % of GDP) | Eurostat |

**Source:** Own processing.

**Annex 2.** *Gini coefficient interpolated data*

| Country | Interpolated data for Gini coefficient (years) | Gini coefficient values |
|---|---|---|
| Austria | 2002 | 25.70 |
| Belgium | 2002 | 28.15 |
| Denmark | 1996, 1998, 2000, 2002 | 20.00, 20.50, 21.50, 23.40 |
| Finland | 2003 | 25.75 |
| France | 2003 | 27.60 |
| Germany | 2002, 2003, 2004 | 25.28, 25.55, 25.83 |
| Greece | 2002 | 33.85 |
| Ireland | 2002 | 29.80 |
| Italy | 2002, 2003 | 30.30, 31.60 |
| Luxembourg | 2002 | 27.30 |
| Netherlands | 2003, 2004 | 26.97, 27.93 |
| Portugal | 2002, 2003 | 37.27, 37.53 |
| Spain | 2003 | 31.00 |
| Sweden | 2003 | 23.00 |
| United Kingdom | 2004 | 34.30 |

**Source:** Own calculations using Eviews 9.0

**Annex 3.** *Domestic credit to private sector (% of GDP) interpolated data*

| Country | Interpolated data for credit to private sector (years) | Credit to private sector (% of GDP) values |
|---|---|---|
| Austria | 1998, 1999, 2000 | 97.61, 95.04,92.46 |
| Belgium | 1998, 1999, 2000 | 72.01, 69.99, 67.96 |
| Denmark | No data missing | No data missing |
| Finland | 1999, 2000 | 51.14, 51.93 |
| France | 1998, 1999, 2000 | 79.02, 78.23, 77.45 |
| Germany | 1999, 2000 | 112.87, 112.46 |
| Greece | No data missing | No data missing |
| Ireland | 1999, 2000 | 80.90, 76.39 |
| Italy | 1999, 2000 | 57.32, 58.96 |
| Luxembourg | 1998, 1999, 2000 | 86.03, 83.91, 81.77 |
| Netherlands | 1998, 1999, 2000 | 101.47, 104.87, 108.27 |
| Portugal | 1999, 2000 | 97.27, 106.12 |
| Spain | 1999, 2000 | 86.99, 91.06 |
| Sweden | No data missing | No data missing |
| United Kingdom | No data missing | No data missing |

**Source:** Own calculations using Eviews 9.0

**Annex 4.** *Model estimation*

Dependent Variable: D(GINI)
Method: Panel EGLS (Cross-section SUR)
Sample (adjusted): 1997 2014
Periods included: 18
Cross-sections included: 15
Total panel (balanced) observations: 270
Linear estimation after one-step weighting matrix

| Variable | Coefficient | Std. Error | t-Statistic | Prob. |
|---|---|---|---|---|
| D(GINI(-1)) | 0.898591 | 0.051850 | 17.33047 | 0.0000 |
| D(CREDITP(-1)) | 0.004989 | 0.001123 | 4.441531 | 0.0000 |
| D(HE(-1)) | -0.104071 | 0.034091 | -3.052751 | 0.0025 |
| D(ED(-1)) | -0.139070 | 0.057220 | -2.430437 | 0.0158 |
| D(UN(-1)) | 0.062590 | 0.013065 | 4.790733 | 0.0000 |
| D(TAX(-1)) | 0.044759 | 0.012317 | 3.633950 | 0.0003 |
| D(CPI) | -0.028655 | 0.004364 | -6.566117 | 0.0000 |
| D(OPENNESS) | -0.019471 | 0.001453 | -13.40060 | 0.0000 |
| D(GFCF) | 0.081449 | 0.010434 | 7.806441 | 0.0000 |
| C | 0.060178 | 0.012139 | 4.957532 | 0.0000 |
| UT(-1) | -1.121507 | 0.063090 | -17.77616 | 0.0000 |

Weighted Statistics

| | | | |
|---|---|---|---|
| R-squared | 0.809313 | Mean dependent var | 0.217654 |
| Adjusted R-squared | 0.801951 | S.D. dependent var | 2.283226 |
| S.E. of regression | 1.000950 | Sum squared resid | 259.4922 |
| F-statistic | 109.9247 | Durbin-Watson stat | 1.893735 |
| Prob(F-statistic) | 0.000000 | | |

**Source:** Own calculations using Eviews 9.0